# Influence of the pre-treatment anneal on Co–germanide Schottky contacts

L. Lajaunie [a],[x], M.L. David [a], F. Pailloux [a], C. Tromas [a], E. Simoen [b], C. Claeys [b,c], J.F. Barbot [a]

[a] PHYMAT, Université de Poitiers/CNRS, UMR 6630, SMP2MI, Bd Marie et Pierre Curie, BP 30179, 86 962 Futuroscope-Chasseneuil Cedex, France
[b] IMEC, Kapeldreef 75, B-3001 Leuven, Belgium
[c] EE Department, KU Leuven, Kasteelpark Arenberg 10, B-3001 Leuven, Belgium

## ABSTRACT

A thin cobalt layer is deposited by electron beam evaporation onto a germanium substrate after an *in situ* cleaning annealing at 400 or 700 °C. The effect of these pre-treatments on the Co/Ge Schottky barrier properties and on the germanide formation is investigated by using different techniques. A strong influence of the pre-treatment is observed. The pre-treatment at 700 °C removes the native oxide but enhances the diffusion of contaminants. After post-metal deposition annealing, the sample pre-treated at 700 °C shows a double layer structure due to interdiffusion, whereas some large isolated islands are present in the sample pre-treated at 400 °C.

## 1. Introduction

Due to extreme scaling, Si-based transistor technology is approaching its physical limits. To follow the required improvement of device performances, new approaches have to be developed. The use of germanium as channel material attracts much attention because of the higher mobility of electrons and holes in germanium than in silicon and its good compatibility with high-k materials [1]. Like self-aligned silicides in Si-based CMOS technology, germanides are considered in Ge-based CMOS to minimize the sheet resistance and to achieve low contact resistances on gate source and drain areas. Among those germanides, Co-germanides seem to be good candidates as they can be formed at low temperatures (400–600 °C) without any metal diffusion into the substrate [2]. However, few studies have been carried out concerning the microstructure and the morphology of the Co–germanide films. One of our major concerns is the influence of the substrate preparation on the formation and on the electrical properties of the Co–germanide films. It has been shown that the nature of the crystallinity of germanium substrate strongly influences the solid-state reaction in the Co/Ge system [3]. Moreover, electrical active defects that can modify the electrical properties of the Schottky contact have been found to emerge depending on the deposition technique [4]. The substrate preparation is always a tricky problem. In most of the previous studies, prior to deposition, the germanium substrate is dipped in a diluted HF solution to remove the Ge native oxide. However, some authors claimed that this sample cleaning is not sufficient and report that some amount of germanium oxide remains after such a pre-treatment [5]. In this work the main objective is to study the impact of the pre-treatments, called "cleaning" pre-annealing, on the electrical characteristics of the devices and on the germanide formation.

## 2. Experimental

N-type (1 0 0) germanium substrates ([Sb]~$1 \times 10^{14}$ at/cm$^3$) were placed in the deposition chamber under a vacuum of $10^{-8}$ Torr. Prior to the deposition, they were annealed at either 400 or 700 °C during 60 min. After cooling down to room temperature during several hours, a 22-nm-thick (measured by reflectivity) cobalt film was deposited by electron beam evaporation. Then, immediately after deposition a second annealing was performed (700 °C, 40 min) *in situ* to form the germanide films. This thermal budget has been chosen to introduce a sufficient concentration of cobalt into the substrate, allowing a study of the cobalt-related trap by deep-level transient spectroscopy (DLTS) [6].

[x] Corresponding author.
*E-mail address:* luc.lajaunie@etu.univ-poitiers.fr (L. Lajaunie).

The electrical characterisation of the diodes and DLTS measurements were performed using a DL8000 Biorad apparatus using a He cryostat. For this purpose, a deposition through a 2-mm diameter mask was performed simultaneously as the thin-film deposition. The thin films of cobalt and germanides were studied by both conventional and high-resolution transmission electron microscopy (HRTEM) using a JEOL 3010 (300 kV, $LaB_6$, point-to-point resolution = 0.19 nm) and a JEOL 2200 FS (200 kV, FEG, in-column O-filter, point-to-point resolution = 0.23 nm), respectively. The TEM samples were prepared in the cross-section geometry using mechanical polishing down to 10 nm and then ion milled in a GATAN-PIPS apparatus at low energy (2.5 keV) and small angles (+-7°). To get a complete set of information about the germanide films, the annealed samples were also characterized by atomic force microscopy (AFM), scanning electron microscopy (SEM) and energy-dispersive X-ray spectroscopy (EDX) techniques.

## 3. Results and discussion

### 3.1. As-deposited samples

Fig. 1 displays bright-field TEM pictures of the cobalt thin films after deposition in the cross-section geometry (X-TEM). To illustrate the effect of the cleaning procedure, a reference sample without any pre-treatment before the cobalt deposition is also imaged (Fig. 1a). All the samples show a continuous thin layer of cobalt. The thickness of this layer is in the range 22–25 nm in good agreement with the reflectivity measurements. At the interface between the cobalt layer and the germanium substrate, a thin amorphous-like layer, 2 nm thick, is observed in both the reference sample and the sample pre-treated at 400 1C. No evidence of this amorphous layer is seen in the picture corresponding with the sample cleaned at 700 °C. This result shows that a threshold temperature $T_s$, 400<$T_s$<700 °C, exists, above which the native oxide is removed under high vacuum. This temperature is, however, strongly dependent on the atmosphere [7]. The diffraction pattern (not shown here) of the untreated sample displays rings ascribed to the nanocrystalline hexagonal-closed-pack cobalt phase. The indexation of all the diffraction patterns is still under investigation; nevertheless since no differences are seen by X-ray diffraction experiments (not shown here) between the samples, the cleaning procedure is supposed to bring no or few effect on the microstructure of the cobalt film.

Table 1 shows the doping density and the Schottky barrier height extracted from the *C–V* characteristics at 270 K. No significant modification of the doping density caused by the pre-treatment is observed. A strong increase in the barrier height is, however, observed after pre-treatments. These values are higher than the one reported after the formation of the germanide films after a post-deposition annealing [8]. It should be noted that during the DLTS measurements, the leakage currents decrease with the increase in the temperature of the pre-treatment. Similar improvements of the electrical characteristics were reported after cleaning annealing on Ge MOSFET [7].

Table 1
Free electron concentrations and Schottky barrier heights extracted from the *C–V* characteristics at 270 K

| Sample | $n$ (cm$^{-3}$) | $j_B$ (eV) |
|---|---|---|
| No pre-treatment | $1.1 \times 10^{14}$ | 0.33 |
| Pre-treatment 400 1C, 60 min | $9.9 \times 10^{13}$ | 0.45 |
| Pre-treatment 700 1C, 60 min | $1.1 \times 10^{14}$ | 0.49 |

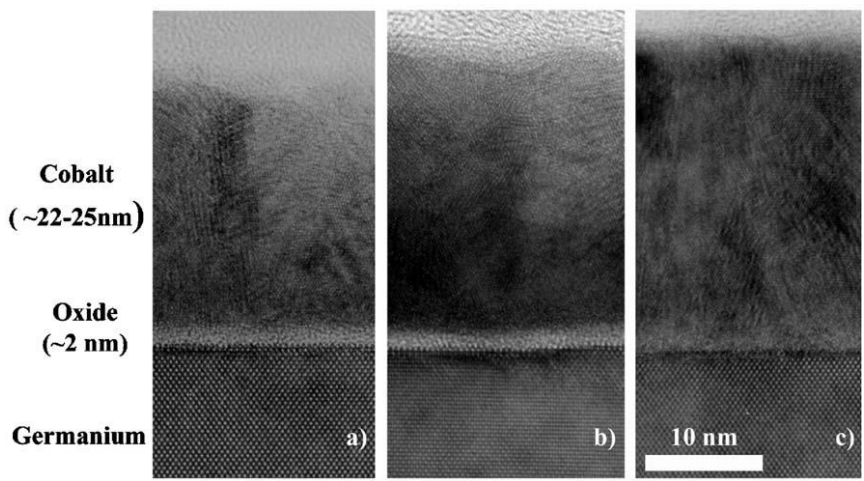

Fig. 1. TEM bright-field micrographs of the cobalt thin films on germanium substrate deposited by e-beam evaporation: (a) reference sample, (b) pre-treated sample at 400 °C and (c) pre-treated sample at 700 °C.

Fig. 2 shows the DLTS spectra recorded on the reference sample and on the two different pre-treated samples. As seen, the reference sample shows three electron traps $E_5$, $E_6$ and $E_7$. Whereas no traps are observed after the 400 °C pre-treatment, several electron traps $E_1$, $E_2$, $E_3$, $E_4$ and $E_7$ and one hole trap $H_1$ are revealed after the 700 °C pre-treatment. The characteristics of these lines are summarized in Table 2. Due to the small amplitude of the lines and the overlap of some peaks, it has not been possible to precisely determine all the signatures. However, the $E_8$ line has been assigned to the $Cu_s^{2-/3-}$ level by comparing with the DLTS spectrum of a Ge sample contaminated with Cu. Copper is a common contaminant of germanium and its presence has been reported before [9,10]. The origin of such contamination is still unclear; however, it has been proposed that co-sputtering of Cu during the metal deposition may be at the origin of the contamination [6].

According to the literature, specifics traps are reported in germanium after Pd e-beam evaporation [4]. However, both $E_5$ and $E_6$ do not correspond with such traps. Moreover, since $E_5$ is not observed after the two pre-treatments, it is unlikely that $E_5$ is introduced by the e-beam evaporation. The same holds for $E_6$. This suggests that the $E_5$ and $E_6$ traps are present in the as-grown sample and that these traps disappear during the pre-treatment. As for the traps observed after the 700 °C pre-treatment, they must be associated with metal contamination during the post-metal deposition annealing. Indeed, these traps show similar signatures than previously reported metal-related traps. For instance, $E_2$, $E_3$, $E_4$ and $H_1$ could be, respectively, assigned to $Au_s^{2-/3-}$, Ni complex, $Au_s^{-/2-}$ and $Cu_s^{-/2}$ [11,12]. This shows that a temperature of 400 °C is not sufficient for the diffusion of such contaminants into the germanium substrate and/or

Table 2
Concentrations and signatures of the deep levels observed after cobalt deposition on the different samples

| Deep levels | $E_C-E_T$ (eV) | s (cm$^2$) | $N_T$ (cm$^{-3}$) | Sample |
| --- | --- | --- | --- | --- |
| $E_1$ | – | – | – | Pre-treated 700 °C |
| $E_2$ | 0.07 | $3 \times 10^{-15}$ | $1.6 \times 10^{12}$ | Pre-treated 700 °C |
| $E_3$ | 0.18 | $2 \times 10^{-13}$ | $3 \times 10^{11}$ | Pre-treated 700 °C |
| $E_4$ | 0.22 | $2 \times 10^{-15}$ | $1.8 \times 10^{11}$ | Pre-treated 700 °C |
| $E_5$ | 0.21 | $2 \times 10^{-16}$ | $5 \times 10^{11}$ | Reference |
| $E_6$ | 0.31 | $8 \times 10^{-15}$ | $1.1 \times 10^{11}$ | Reference |
| $E_7$ | 0.34 | $2 \times 10^{-15}$ | $1 \times 10^{11}$ | Reference and 700 °C |
| $H_1$ | 0.31 | $6 \times 10^{-14}$ | - | Pre-treated 700 °C |

The energy level of the hole trap $H_1$ is referred to the valence band and its signature has been determined under pulse injection. Due to small concentrations and overlapping it has not always been possible to derive any signature.

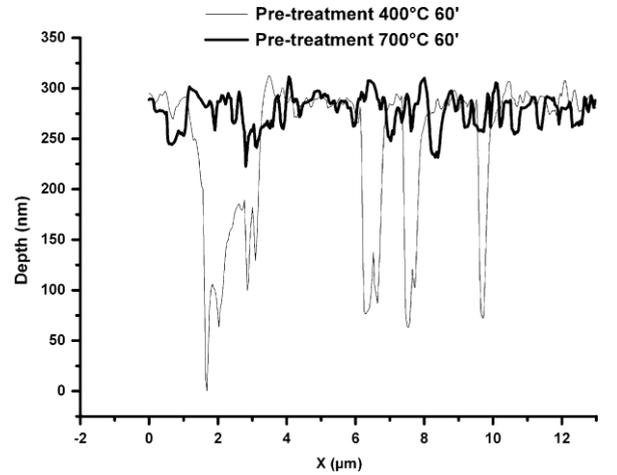

Fig. 3. AFM surface profiles of the samples pre-treated at 400 and 700 °C after post-metal annealing at 700 °C.

that the oxide layer at the Co/Ge interface of the 400 °C pre-treated sample acts as a barrier for the diffusion of the contaminants into the substrate.

### 3.2. Post-metal deposition annealing

Fig. 3 displays the scans of the surface extracted from AFM pictures after the post-metal deposition annealing. As seen, the topography of the surface strongly depends on the pre-treatment. For the sample pre-treated at 400 °C the surface is very rough, with some height differences up to 300 nm. The surface of the sample pre-treated at 700 °C is smoother, with height differences in the range of 50 nm. This shows that the presence of the oxide at the Co/Ge interface strongly influences the formation mechanisms of germanides. SEM and TEM experiments were performed on both samples to get a better insight of the morphology. Fig. 4 displays an SEM picture in the cross-section geometry of the sample pre-treated at 400 °C and annealed. Several isolated islands are clearly visible at the surface, which explains the large differences (bumps) revealed by the AFM profile. According to the AFM profile, these islands are of various sizes, with heights ranging from 50 to 300 nm. The chemical analysis (EDX, not shown here) shows that the Ge/Co ratio is smaller in the large bumps than in the zones in between. On the contrary, no significant difference in the Ge/Co ratio has been revealed by the EDX spectra performed at different areas onto the surface of the sample pre-treated at 700 ℃. The formation of such large scattered bumps is not well understood and still under investigation. Fig. 5a shows a low-magnification X-TEM picture of the sample pre-treated at 700 °C and annealed. Two distinct layers are formed above the germanium substrate. The top layer is discontinuous, with a thickness up to 35 nm. The discontinuity of this layer explains the AFM profiles observed previously. Beneath the top layer, the underlayer contains several grains with a thickness from 40 to 70 nm. On the contrary whereas the interface between the top layer and the underlayer is smooth, the interface between the underlayer and the germanium substrate is rough. Indexing of diffraction patterns together with fast Fourier transform analysis of the HRTEM picture (Fig. 5b) reveals that the top layer is, on its continuity, a single crystal of $Co_5Ge_7$ seen along the [10 0] direction. Similar analysis reveals that the underlayer is formed of grains of $Co_5Ge_7$ and $CoGe_2$ seen, respectively, along the [10 0] and the [0 11] directions. The above results indicate that not only the top layer is epitaxially aligned to the underlayer, but also the underlayer epitaxially grows on the germanium substrate. The epitaxial orientations are [10 0] (0 0 1) $Co_5Ge_7$∥[0 11] (0 0 1) $CoGe_2$∥[110] (0 0 1) Ge and [10 0] (0 0 1) $Co_5Ge_7$∥[10 0] (0 0 1) $Co_5Ge_7$∥[110] (0 0 1) Ge. Formation of such a double layer structure reveals the interdiffusion of cobalt and germanium. No similar microstructure has been found in the literature yet. This may relate to the duration of the post-metal deposition (40 min) annealing, which is larger than those commonly used in the metallization processes (RTA).

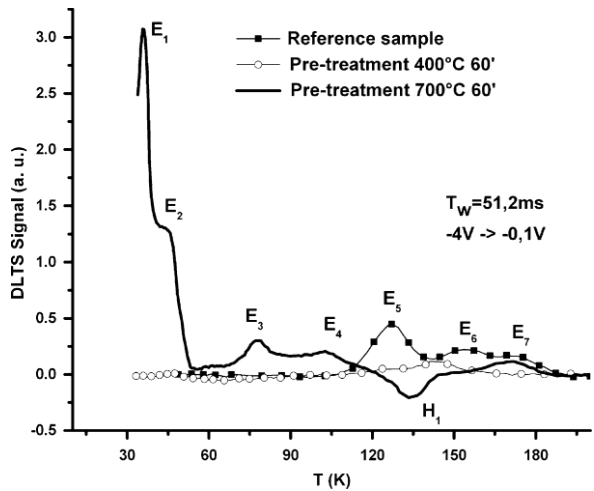

Fig. 2. DLTS spectra of the Co/Ge Schottky barriers after metal deposition. $T_W$ = 51.2 ms, $T_P$ = 5 ms, $V_R$ = −4 V, $V_P$ = −0.1 V.

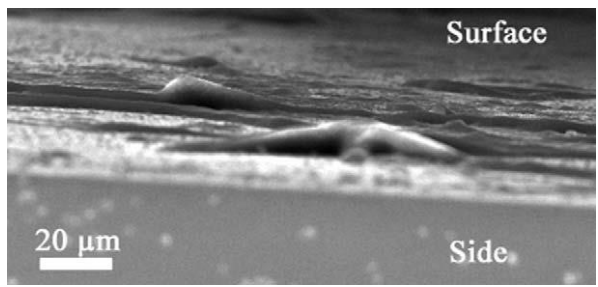

Fig. 4. SEM image of the sample pre-treated at 400 °C after post-metal annealing at 700 °C in cross-section geometry.

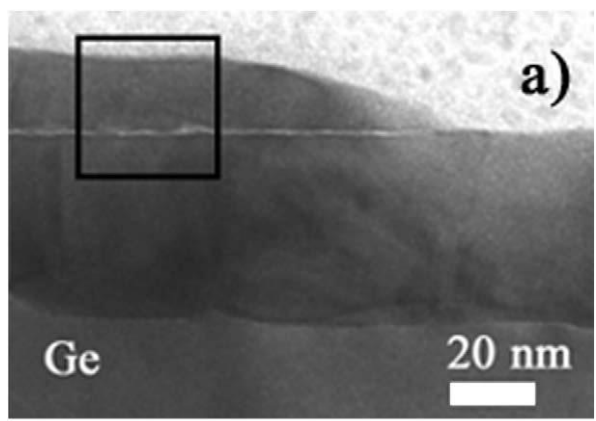

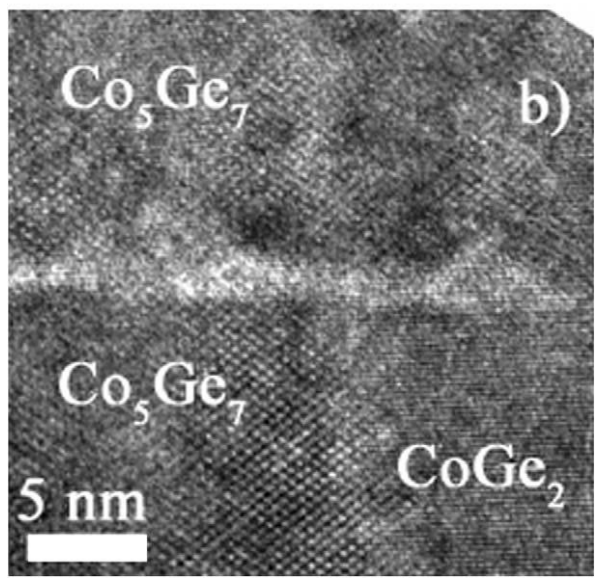

Fig. 5. (a) X-TEM micrographs of the sample pre-treated at 700 °C after post-metal deposition annealing: (a) at low magnification and (b) HTREM micrograph.

4. Conclusion

The effects of cleaning annealing before cobalt deposition on the Co/Ge Schottky barrier properties and on the cobalt germanide formation have been investigated. A temperature higher than 400 °C is necessary to remove the native oxide. Electrical analyses reveal an increase of the Schottky barrier height and a decrease of the leakage currents according to the temperature of the pre-treatments. However, during the annealing necessary to remove the native oxide, the unavoidable diffusion of contaminants occurs. After a pre- and post-treatment at 700 °C, the samples show two layers of germanides due to interdiffusion. The top layer is a discontinuous monocrystal of $Co_5Ge_7$, whereas the underlayer is made of grains of $Co_5Ge_7$ and $CoGe_2$. The epitaxial orientations have been clearly established. In the case of the pre-treatment at 400 °C, a post-metal deposition annealing leads to the formation of large isolated islands whose mechanisms of formation are not clear yet.